\documentclass[superscriptaddress,amsmath,amssymb,prd,preprintnumbers,showpacs,twocolumn,nofootinbib]{revtex4-2}

\usepackage[colorlinks=true,pdfstartview=FitV,linkcolor=blue,citecolor=blue,urlcolor=blue,breaklinks=true]{hyperref}
\usepackage{graphicx}
\usepackage[T1]{fontenc} % if needed
\usepackage{amssymb,amsmath,bm,natbib}
\usepackage{color}
\usepackage{slashed}
\usepackage{graphics}
\usepackage{graphicx}
\usepackage[utf8]{inputenc}
\usepackage[caption=false]{subfig}
\usepackage{hyperref}
\usepackage{url}
\usepackage{dsfont}
\usepackage{float}
\usepackage{cancel}
\usepackage{units}
\usepackage{blindtext}
\usepackage[utf8]{inputenc}
\usepackage{upgreek}
\usepackage{booktabs}
\usepackage[dvipsnames,table,xcdraw]{xcolor}
\usepackage{enumerate}
\usepackage{mathtools}
\usepackage{soul,color}
\usepackage[normalem]{ulem}
\newcommand{\orcid}[1]{\href{https://orcid.org/#1}{\includegraphics[width=10pt]{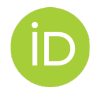}}}

\begin{document}

\title{Reduced geometry and its role in explicit spacetime symmetry violation}

\author{Carlos M. Reyes\orcid{0000-0001-5140-6658}}
\email{creyes@ubiobio.cl}
\affiliation{Centro de Ciencias Exactas, Universidad del B\'{i}o-B\'{i}o, Chill\'{a}n, Casilla 447, Chile}

\author{C\'esar Riquelme\orcid{0000-0003-0837-3891}}
\email{ceriquelme@udec.cl}
\affiliation{Departamento de F\'{i}sica, Universidad de Concepci\'{o}n, Concepci\'on, Casilla 160-C, Chile}

\author{Marco Schreck\orcid{0000-0001-6585-4144}}
\email{marco.schreck@ufma.br}
\affiliation{Departamento de F\'{i}sica, Universidade Federal do Maranh\~{a}o, Campus Universit\'{a}rio do Bacanga, S\~{a}o Lu\'{i}s (MA), 65085-580, Brazil}

\author{Alex Soto\orcid{0000-0002-9731-5521}}
\email{arsoto1@uc.cl}
\affiliation{School of Mathematics, Statistics and Physics, Newcastle University, Newcastle upon Tyne, NE1 7RU, UK}

\begin{abstract}

The incompatibility of explicit diffeomorphism violation with Riemannian geometry within the gravitational Standard-Model Extension (SME) is revisited. We review two methods of how to deal with this problem. The first is based on an approach proposed originally by St\"{u}ckelberg and the latter is to restrict spacetime geometry via the dynamical field equations and the second Bianchi identities. Moreover, a third technique is introduced
in this work, which relies on isometries of a gravitational system. Our conclusion is that an SME background field configuration compatible with Riemannian geometry is more likely to be determined the more diffeomorphisms are isometries of the particular system. The proposal is demonstrated to work for cosmological time evolution with the SME backgrounds $u$ and $s^{\mu\nu}$ present. This finding has the potential to provide an alternative treatment of explicit spacetime symmetry violation in gravity.

\end{abstract}

\maketitle
\section{Introduction}
\label{sec:introduction}

General Relativity (GR) is intrinsically nonlinear, which makes finding solutions of the Einstein equations a highly challenging task. Therefore, it turns out to be valuable to impose symmetries before solving the dynamical field equations for a gravitational system under consideration. In this context, Noether symmetries are of paramount interest. They arise when the variation of the action of a field theory vanishes under certain infinitesimal transformations. Invariances under diffeomorphisms and gauge transformations are prominent examples for Noether symmetries.

Based on that, the literature distinguishes between Noether symmetries generated by Killing vector fields and those admitting conformal Killing vector fields; see, e.g., Ref.~\cite{Blau:2020}. The former are related to isometries, i.e., transformations leaving the metric form-invariant. The latter leave the metric form-invariant except for an overall Weyl rescaling. We also note in passing that the equations of motion can exhibit symmetries that the action does not have. These are known as Lie symmetries, but they will not be of significance in our paper.

In the following, we focus on spacetime symmetry violation, in particular, violations of diffeomorphism invariance, which is one of the most remarkable properties of GR. Whenever we speak of diffeomorphism violation, we actually refer to particle diffeomorphisms~\cite{Bluhm:2014oua}, whereas the theory is understood to be invariant under general coordinate transformations (observer diffeomorphisms~\cite{Bluhm:2014oua}).

In general, there are two possibilities of how to violate a symmetry, which is by either spontaneous or explicit means. The first rests upon tensor-valued fields acquiring symmetry-violating vacuum expectation values dynamically. Such mechanisms were already proposed in bosonic string field theory at the beginning of the 90s \cite{Kostelecky:1988zi,Kostelecky:1989jp,Kostelecky:1989jw,Kostelecky:1991ak,Kostelecky:1994rn}. Spontaneous diffeomorphism violation has been the preferred option for spacetime symmetry violation in the gravitational Standard-Model Extension (SME) \cite{Kostelecky:2003fs,Kostelecky:2020hbb} for two decades since it was demonstrated to be dynamically consistent with Riemannian geometry \cite{Bluhm:2004ep,Bluhm:2007bd}. 

In contrast, explicit diffeomorphism violation is plagued by conflicts between the dynamics and Riemannian geometry. Reconciling the modified Einstein equations with the contracted second Bianchi identities for the Riemann tensor implies a number of coupled partial differential equations for the SME background coefficients, which are challenging to solve. This important finding has been coined the no-go result in the contemporary literature~\cite{Bluhm:2014oua,Kostelecky:2003fs,Kostelecky:2020hbb,Bluhm:2016dzm} and it has basically led to two possible interpretations. A nondynamical SME background field is either restricted severely when Riemannian geometry is to be maintained or one is forced to work in a beyond-Riemannian setting such as Finsler geometry~\cite{Kostelecky:2010hs,Kostelecky:2011qz,AlanKostelecky:2012yjr,Schreck:2015seb}.

Until now two approaches of how to tackle this problem while maintaining Riemannian geometry have been presented in the literature. The first is to use a procedure that is known as the Stückelberg trick. Stückelberg originally introduced an auxiliary scalar field into Proca theory \cite{Stueckelberg:1938,Ruegg:2003ps} to restore gauge symmetry such that there is a smooth limit when the Proca mass approaches zero. This idea was taken over to massive gravity \cite{Arkani-Hamed:2002bjr,Hinterbichler:2011tt} as well as to the gravitational SME with explicit diffeomorphism breaking~\cite{Bluhm:2019ato}, where it works at the level of linearized gravity. Here, the technique rests upon several auxiliary fields that mimic the Nambu-Goldstone modes arising in a setting of spontaneous spacetime symmetry violation. Therefore, the minimum number of excitations is reintroduced to restore the broken symmetries. By doing so, at least some of the essential properties of a setting of spontaneous spacetime symmetry breaking are recovered, although the background field is nondynamical. However, the approach does not reintroduce the Higgs-like modes, which also arise naturally when symmetry breaking is spontaneous.

The second method is to restrict spacetime geometry to suppress diffeomorphism violation dynamically. In particular, this approach was applied by Jackiw and Pi to the gravitational Chern-Simons (CS) term in four spacetime dimensions \cite{Jackiw:2003pm}. The latter can be expressed as a divergence of the CS topological current, which, after suitable integrations by parts, is identified with the Chern-Pontryagin scalar density $2{}^{*}R^{\sigma\phantom{\tau}\mu\nu}_{\phantom{\sigma}\tau}R^{\tau}_{\phantom{\tau}\sigma\mu\nu}=:2{}^{*}RR$, where $R^{\tau}_{\phantom{\tau}\sigma\mu\nu}$ is the Riemann curvature tensor and ${}^{*}R^{\tau\phantom{\sigma}\mu\nu}_{\phantom{\tau}\sigma}:=(1/2)\varepsilon^{\mu\nu\alpha\beta}R^{\tau}_{\phantom{\tau}\sigma\alpha\beta}$ is its dual. Using a suitable normalization and integrating the latter over spacetime provides a $\mathbb{Z}_2$ topological quantity, which is known as the second Chern number or the gravitational instanton number~\cite{Obukhov:1995eq}. Now, for the CS-like term to be consistent with the contracted second Bianchi identities, one must impose that ${}^{*}RR=0$, which reduces the space of all possible geometries to those obeying this requirement. Consequently, the topological properties characterized by the second Chern number must be trivial for spacetimes to be dynamically consistent in this case.

In their recent Ref.~\cite{Bailey:2024zgr}, Bailey \textit{et al.} presented another example of how to reduce 
spacetime geometry for a dynamically consistent theory with nondynamical background fields. They study the 
gravitational sector of the minimal SME with tensor-valued background fields $u$, $s^{\mu\nu}$, and 
$t^{\mu\nu\varrho\sigma}$ violating diffeomorphism invariance explicitly. For simplicity, let us be content with the scalar $u$, the trace $s^{\alpha}_{\phantom{\alpha}\alpha}:=g^{\alpha\beta}s_{\alpha\beta}$ of the $s$ coefficients, and the double trace $t^{\alpha\beta}_{\phantom{\alpha\beta}\alpha\beta}:=g^{\alpha\varrho}g^{\beta\sigma}t_{\varrho\sigma\alpha\beta}$ of the $t$ coefficients, where the latter two are hybrid coefficients involving the dynamical (inverse) spacetime metric $g^{\mu\nu}$. By taking into account the second Bianchi identities of Riemannian geometry, three requirements are derived to be satisfied by the background fields. Disregarding the possibility of mutual cancelations between distinct terms, these conditions can be written in the generic form
\begin{align}
\label{eq:conditions-bailey}
0=R\nabla^{\nu}f(u,s^{\alpha}_{\phantom{\alpha}\alpha},t^{\alpha\beta}_{\phantom{\alpha\beta}\alpha\beta})-\nabla_{\mu}g(s^{\mu\nu}R,t^{\mu\alpha\nu}_{\phantom{\mu\alpha\nu}\alpha}R)\,,
\end{align}
with suitable functions $f$ and $g$. The remarkable observation about Eq.~\eqref{eq:conditions-bailey} is that every term depends on the Ricci scalar $R$, i.e., each condition is interpreted to reduce all spacetime geometries to the subset characterized by $R=0$. In particular, for the $u$ sector, it must hold that $R\nabla^{\nu}u=0$. Interestingly, the latter was already pointed out in a general argument in Ref.~\cite{Jackiw:2003pm}, even before the minimal gravitational SME~\cite{Kostelecky:2003fs} was introduced to the community. So spacetimes endowed with certain nondynamical scalar background fields are only then dynamically consistent if their scalar curvature vanishes.

Another approach has been applied to a cosmology modified by the $t$ sector of the minimal gravitational SME with explicit symmetry violation~\cite{Reyes:2024hqi}. Isotropy and homogeneity of spacetime, which many cosmological models are based on, were imposed on the modified-gravity theory considered. This was accomplished by ensuring two essential properties of the purely spacelike part of the second-rank tensor that contains the background field and modifies the Einstein equations. It was required that the latter be form-invariant under the six isometries generated by their corresponding spatial Killing vector fields and that its symmetric, traceless part vanish. As a result, spacetime geometry is effectively restricted such that it is not in conflict with dynamics. At least, this holds when focusing on diffeomorphism breaking induced by the purely spacelike background coefficients $t^{abcd}$.

We emphasize that reducing geometry is a procedure that is, in a certain sense, related to the dynamics of the modified-gravity theory under study. The modified Einstein equations still have to be solved for a specific system of interest. However, restricting spacetime geometry provides a guidance for proposing a suitable metric \textit{ansatz} such that inconsistencies between dynamics and geometry can be neatly avoided. An example for a spacetime geometry consistent with both ${}^{*}RR=0$ and $R=0$ is based on the Schwarzschild \textit{ansatz} for a static, spherically symmetric gravitational system without the presence of matter energy-momentum.

In Sec.~\ref{sec:killing-isometries} of this paper, we will describe the basic ideas behind the third possibility of how to render a setting of explicit spacetime symmetry violation consistent with the dynamical field equations. This approach was originally motivated by the previously mentioned results of Ref.~\cite{Reyes:2024hqi}. It consists of restricting spacetime geometry by imposing symmetries on the metric according to certain properties of a gravitational system at hand. The forthcoming formulation crucially relies on the concept of Killing vector fields outlined at the beginning. Section~\ref{sec:implementation} then demonstrates the general procedure with examples in a cosmological setting. The paper ends with us concluding on the crucial outcomes in Sec.~\ref{sec:conclusions}.

\section{Killing fields, isometries, and explicit breaking}
\label{sec:killing-isometries}

Let us consider the following generic gravitational action:
\begin{equation}
\label{eq:modified-action}
S=\int\mathrm{d}^4x\,\frac{\sqrt{-g}}{2\kappa}\,\Big[R+\mathcal{L}'(g_{\mu\nu},\bar{k}^{\alpha\beta\dots\omega})\Big]+S_B\,,
\end{equation}
where $\kappa=8\pi G_N$ with Newton's constant $G_N$, the spacetime metric $g_{\mu\nu}$ with determinant $g=\det(g_{\mu\nu})$, and the associated Ricci scalar $R$. The Lagrange density $\sqrt{-g}\mathcal{L}'$ involves the coefficients $\bar{k}^{\alpha\beta\dots\omega}$ of a generic nondynamical background field, which makes the action noninvariant under particle diffeomorphisms. Moreover, for consistency with the stationary-action principle, we include a suitable extension of the Gibbons-Hawking-York (GHY) boundary term~\cite{York:1972sj,Gibbons:1976ue,Reyes:2023sgk,Reyes:2021cpx,Reyes:2022mvm} described by $S_B$. The action is constructed to be invariant under observer diffeomorphisms. Then, the variation of $S$ with respect to such transformations reads:
\begin{subequations}
\label{eq:variation-action-observer-transformation}
\begin{align}\label{eq:variation-action-observer-transformation_a}
\delta S_{\mathrm{obs}}&=\int\mathrm{d}^4x\,\frac{\sqrt{-g}}{2\kappa}\left(-G^{\mu\nu}+ T'^{\mu\nu}\right)\delta g_{\mu\nu} \notag \\
&\phantom{{}={}}+\int\mathrm{d}^4x\,\frac{\sqrt{-g}}{2\kappa}J_{\alpha\beta\dots\omega}\delta \bar{k}^{\alpha\beta\dots\omega}=0\,,
\end{align}
where we defined
\begin{equation}
T'^{\mu\nu}=: \frac{1}{\sqrt{-g}} \frac{\delta(\sqrt{-g}\mathcal{L}')}{\delta g_{\mu\nu}}\,,\quad J_{\alpha\beta\dots\omega}=:\frac{\delta\mathcal{L}'}{\delta\bar{k}^{\alpha\beta\dots\omega}}\,,
\end{equation}
\end{subequations}
and $G_{\mu \nu}=R_{\mu \nu}-(R/2)g_{\mu\nu}$ is the Einstein tensor with the Ricci tensor $R_{\mu\nu}$ and the Ricci scalar $R$. Since the presence of the background field violates particle diffeomorphism invariance, there is a mismatch between the variations of the action with respect to general coordinate transformations and particle diffeomorphisms. The latter can be expressed via the following integral equation:
\begin{equation}
\label{eq:principal-equation}
\delta S_{\mathrm{part}}+\int\mathrm{d}^4x\,\frac{\sqrt{-g}}{2\kappa}J_{\alpha\beta\dots\omega}\delta \bar{k}^{\alpha\beta\dots\omega} =\delta S_{\text{obs}}  \,.
\end{equation}
Thus, Eq.~\eqref{eq:principal-equation} describes the origin of the clash between nondynamical background fields and Riemannian geometry in a succinct manner. 
For a diffeomorphism being an isometry with Killing vector field $\chi$, we demand that the metric be form-invariant under this transformation: 
$\delta g_{\mu\nu}=\mathcal{L}_{\chi}g_{\mu\nu}=0$, where $\mathcal{L}_{\chi}$ denotes the Lie derivative~\cite{Yano:1957,Carroll:1997ar} along $\chi$. Moreover, we specifically use $\delta\bar{k}^{\alpha\beta\dots\omega}=\mathcal{L}_{\chi}\bar{k}^{\alpha\beta\dots\omega}$. Then, Eq.~\eqref{eq:variation-action-observer-transformation_a} implies
\begin{equation}
\label{int_eq}
\delta S_{\mathrm{obs}}=0=\int\mathrm{d}^4x\,\frac{\sqrt{-g}}{2\kappa}J_{\alpha\beta\dots\omega}\mathcal{L}_{\chi}\bar{k}^{\alpha\beta\dots\omega}\,.
\end{equation}
Hence, there only remains the Lie derivative part of the background field. This leads to the requirement
\begin{equation}
\label{eq:isometry-imposed-background}
\mathcal{L}_{\chi}\bar{k}^{\alpha\beta\dots\omega}=0\,,
\end{equation}
such that the isometry is imposed on the background field. For a diffeomorphism that is not an isometry, it holds that $\delta g_{\mu\nu}=\mathcal{L}_{\xi}g_{\mu\nu}=\nabla_{\mu}\xi_{\nu}+\nabla_{\nu}\xi_{\mu}$ and $\delta
\bar{k}^{\alpha\beta\dots\omega}=\mathcal{L}_{\xi}\bar{k}^{\alpha\beta\dots\omega}$, with the Lie 
derivative $\mathcal{L}_{\xi}$ along the generator~$\xi$ of the diffeomorphism. Performing several
integrations by parts in Eq.~\eqref{eq:variation-action-observer-transformation_a} and employing the 
contracted second Bianchi identities $\nabla^{\mu}G_{\mu\nu}=0$, we arrive at the identity
\begin{align}
\label{eq:principal-equation-differential-form}
2\nabla_{\mu}T'^{\mu}_{\phantom{\mu}\,\,\nu}&=J_{\alpha\beta\dots\omega}
\nabla_{\nu}\bar{k}^{\alpha\beta\dots\omega}+\nabla_{\lambda}(J_{\nu\beta\dots\omega}\bar{k}^{\lambda\beta\dots\omega}) \notag \\
&\phantom{{}={}}+\nabla_{\lambda}(J_{\alpha\nu\dots\omega}\bar{k}^{\alpha\lambda\dots\omega})+\dots \notag \\
&\phantom{{}={}}+\nabla_{\lambda}(J_{\alpha\beta\dots\nu}\bar{k}^{\alpha\beta\dots\lambda})\,.
\end{align}
For the theory to be dynamically consistent, a critical requirement is that the following differential equations be satisfied \cite{Kostelecky:2003fs,Kostelecky:2020hbb}:
\begin{equation}
\label{eq:consistency-requirement}
\nabla_{\mu}T'^{\mu}_{\phantom{\mu}\,\,\nu}=0\,.
\end{equation}%%
Note that the latter result has been well-known for around 20 years. However, below, we intend to provide an alternative view on this problem.
The previous equations enable us to understand how the conflict between explicit spacetime symmetry breaking and dynamics can be resolved, at least for particular systems equipped with isometries. Equation~\eqref{eq:principal-equation} clearly shows the conflict in the variation of the action with respect to observer and particle diffeomorphisms in the presence of a nondynamical background field. Each of the 4 equations holds for 1 out of the 4 diffeomorphism generators, which exist in 4 spacetime dimensions.

Note that the GR action, when formulated in terms of the vierbein, is invariant under local $\mathit{SO}(1,3)$ transformations. The same applies to all of its solutions. Therefore, there are six local Lorentz generators. These are of minor importance in the present context, as any breaking of local Lorentz invariance is not directly evident in Eq.~\eqref{eq:modified-action} that is written in terms of the metric. So even in a modified-gravity theory with nondynamical background fields, we cannot speak of local Lorentz violation without resorting to the dynamical vierbein. Upon suitable contraction of the latter with background fields carrying spacetime indices, hybrid backgrounds emerge leading to scenarios beyond the scope of the paper. Hence, we leave possible local Lorentz violation aside.

Isometries can be interpreted as global symmetries, because a vector field generating the corresponding transformation has constant components in a suitably chosen basis. This is a highly specific property, which is not necessarily valid for an arbitrary diffeomorphism. Since isometries are very special diffeomorphisms with furnished properties, they are also Noether symmetries. Examples for isometries are transformations that could be called manifold rotations, as a homage to Ref.~\cite{Kostelecky:2020hbb}, which introduced the concept of a manifold Lorentz transformation. Manifold rotations mimic usual rotations applied to vector and tensor fields in local frames, but they are actually diffeomorphisms whose Killing vectors satisfy the $\mathfrak{so}(3)$ algebra. Therefore, they do not act in local frames, but on points and sets of the spacetime manifold proper. It can be shown that a $d$-dimensional spacetime manifold is capable of having $d(d+1)/2$ Killing vectors at the maximum, which amounts to 10 for $d=4$.

First of all, let us assume that the gravitational system does not exhibit any isometries that can be identified with diffeomorphisms. In this case, we must resort to Eq.~\eqref{eq:consistency-requirement}. For a rather generic metric of complicated form, the latter set of coupled nonlinear partial differential equations is likely to heavily restrict the background field, maybe even to the point that all the coefficients vanish identically. However, there may be a way out for gravitational systems that exhibit at least a single diffeomorphism corresponding to an isometry, as we shall argue below.

By following this line of reasoning, suppose that the gravitational system is characterized by its invariance under a certain diffeomorphism. For example, the Schwarzschild \textit{ansatz}, which describes a static metric, is invariant under infinitesimal time diffeomorphisms. Such a diffeomorphism is then an isometry of the metric, which allows us to identify the generator of this diffeomorphism with a Killing vector field $\chi$. Now, Eq.~\eqref{eq:isometry-imposed-background} applies, whereupon the nondynamical background field must be compatible with the symmetry of the metric. Consistency between dynamics and Riemannian geometry demands that Eq.~\eqref{eq:consistency-requirement} also be valid. The latter is a consistency requirement, whereas the former takes the role of a physical statement on a symmetry that the gravitational system possesses.

The role of the isometry is to restrict both the metric $g_{\mu\nu}$ and the background field $\bar{k}^{\alpha\beta\dots\omega}$. By doing so, the differential equations of Eq.~\eqref{eq:consistency-requirement} are supposed to simplify, which contributes to finding nontrivial solutions $\bar{k}^{\alpha\beta\dots\omega}$ compatible with the dynamics. The presence of each further isometry that can be identified with a diffeomorphism reduces the complexity of $g_{\mu\nu}$ and $\bar{k}^{\alpha\beta\dots\omega}$ even more. Consequently, the system of differential equations to be solved for the coefficients of the background field is expected to be ever more manageable.

For simple background fields to start with such as the coordinate scalar $u$, all these conditions are likely to only permit background fields that are either trivial or have a rather simple dependence on the spacetime coordinates. However, if the number of Lorentz indices in $\bar{k}^{\alpha\beta\dots\omega}$ increases, in particular, for the nonminimal SME, the background field is supposed to have enough independent components such that some nonzero coefficients are expected to survive. Note that the method described in its generality in the present article was successfully applied to the $t^{\mu \nu \alpha \beta}$ coefficients in Ref.~\cite{Reyes:2024hqi} very recently, which showcases a specific example. Further examples will be presented in Sec.~\ref{sec:implementation}. Thus, isometries turn out to be critical in the quest for nondynamical background fields giving rise to a consistent theory.

For a spacetime with $n$ isometries described by Killing vector fields $\chi^{(i)}$, $i=1\dots n$ the following holds: (i) each Killing vector field generates a diffeomorphism that satisfies Eq.~\eqref{eq:isometry-imposed-background} on its own and (ii) the second term in the left-hand side of Eq.~\eqref{eq:principal-equation}, when expressed in terms of the Lie derivative, vanishes identically along these symmetry directions. This implies that both variations are equal, $\delta S_{\mathrm{part}} =\delta S_{\text{obs}}$, and so, diffeomorphism invariance is partially restored along the Killing directions. For generic diffeomorphisms, which are not necessarily isometries, the identity~\eqref{eq:principal-equation-differential-form} should be consulted. Then, the consistency condition of Eq.~\eqref{eq:consistency-requirement} must hold on-shell, imposing additional dynamical constraints on the backgrounds.
\begin{figure}
\centering
\includegraphics[scale=1]{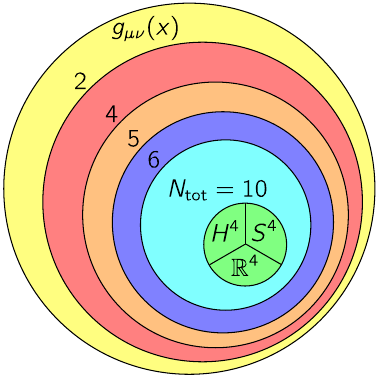}
\caption{Space of all possible spacetime metrics containing metric \textit{ansätze} as well as solutions of the dynamical field equations. This space involves subspaces of metrics with an ever-increasing number of isometries described by $N_{\mathrm{tot}}\in [1,10]$ Killing vector fields. The innermost subspace comprises maximally symmetric spacetimes with $N_{\mathrm{tot}}=10$, where examples are the de-Sitter ($S^4$), Minkowski ($\mathbb{R}^4$), and anti-de-Sitter ($H^4$) metrics. Examples for $N_{\mathrm{tot}}=6$ are the FLRW-type metrics, for $N_{\mathrm{tot}}=5$ the metric of the G\"{o}del Universe, for $N_{\mathrm{tot}}=4$ the Schwarzschild metric, and for $N_{\mathrm{tot}}=2$ the Kerr metric.}
\label{fig:visualization-reduced-geometry}
\end{figure}%%

Maximally symmetric spacetimes with 10 isometries provide a special case. The metric then takes a diagonal form with all its components depending on spacetime coordinates in a simple way. The well-known three classes of these geometries are de-Sitter ($S^4$), Minkowski ($\mathbb{R}^4$), and anti-de-Sitter ($H^4$) spacetimes.  For example, take $g_{\mu\nu}=\eta_{\mu\nu}$, which is the Minkowski metric. The latter has 10 isometries, where 4 of these are identified with translations (diffeomorphisms). By working in Cartesian coordinates, the associated Killing vector fields $\chi^{(i)}$ for $i=1\dots 4$ can be chosen to have constant coefficients. Then, Eq.~\eqref{eq:isometry-imposed-background} definitely holds for constant background field coefficients. This is a convenient property of the nongravitational SME, which is free of any geometrical inconsistencies with dynamics.

Moreover, the presence of isometries can be interpreted as reducing spacetime geometry in an analogous way as it occurs for the gravitational CS term~\cite{Jackiw:2003pm} and the treatment of the minimal gravitational SME in Ref.~\cite{Bailey:2024zgr}, as explained in Sec.~\ref{sec:introduction}. To understand this, we start from the space of all possible spacetime metrics $g_{\mu\nu}$ without any symmetries imposed, which contains all real and symmetric $(4\times 4)$ matrices depending on the spacetime coordinates. Let $\chi^{(i)}$ be a Killing vector for $i=1\dots N_{\mathrm{tot}}$ where $1\leq N_{\mathrm{tot}}\leq 10$ such that $N_{\mathrm{tot}}=10$ matches the maximum number of isometries possible in $d=4$.

For each isometry that exists, the initial space of all metrics is reduced to a subspace, which involves only the metrics in agreement with the isometry required. Increasing the number of Killing vectors one by one, steadily increases the symmetry of the metric and reduces the number of its independent degrees of freedom (see Fig.~\ref{fig:visualization-reduced-geometry}).

%...................................
\section{Implementation}
\label{sec:implementation}
%...................................

In this section, we apply the preceding theoretical framework to specific models exhibiting diffeomorphism symmetry breaking. The \textit{Mathematica} packages \textit{xTensor} and \textit{xCoba}~\cite{xTensor:2024} have served as computational support to crosscheck certain results. Through concrete examples in cosmological time evolution, we demonstrate how geometries with global symmetries --- characterized by Killing vectors --- can maintain the symmetries of the original theory while satisfying Eq.~\eqref{eq:consistency-requirement} as an implication of the second Bianchi identities. Specifically, we focus on explicit breaking arising from  the $u$ and $s^{\mu \nu}$ sectors of the gravitational SME~\cite{Kostelecky:2003fs}, described by the action
\begin{subequations}
\label{eq:explicit_diffeo}
\begin{align}
\label{eq:explicit_diffeo-1}
S_{g}&=\int_{\mathcal{M}} \mathrm{d}^4x\,\frac{\sqrt{-g}}{2\kappa}({R}+\mathcal L'_{u,s})+S_B\,,
\end{align}
where
\begin{align}
\mathcal L'_{u,s}&=-u {R}+s^{\mu \nu}{R}_{\mu \nu}\,.
\end{align}
\end{subequations}
On the four-dimensional spacetime manifold $\mathcal M$, we denote the Ricci tensor by $R_{\mu \nu}$ and the scalar curvature by $R$. The last contribution in Eq.~\eqref{eq:explicit_diffeo-1} is an extended GHY boundary term, which allows us to implement the usual variational principle~\cite{York:1972sj,Gibbons:1976ue,Reyes:2023sgk,Reyes:2021cpx,Reyes:2022mvm}. The tensor-valued background field $s^{\mu \nu}=s^{\mu\nu}(x)$ inherits the symmetries of the Ricci tensor and $u=u(x)$ is a coordinate scalar.

Recall the Friedmann-Lema\^{i}tre-Robertson-Walker (FLRW) metric
\begin{subequations}
\label{metricFLRW}
\begin{align}
   \mathrm{d}s^2=-\mathrm{d}t^2 + a^2(t) \left( \frac{\mathrm{d}r^2}{1-kr^2}
    +r^2 \mathrm{d}\Omega^2  \right)  \,,
\end{align}
with
\begin{align}
  \mathrm{d}\Omega^2= \mathrm{d}\theta^2+\sin^2\theta \mathrm{d}\phi^2      \,,
\end{align}
\end{subequations}
written in terms of three-dimensional spherical coordinates $(r,\theta,\phi)$. We introduced the cosmic scale factor $a(t)$ and the spatial curvature $k$, which gives rise to a closed ($k=1$), flat ($k=0$) or open ($k=-1$) universe.

%...................................
\subsection{$u$ sector}
%...................................

Each sector of Eq.~\eqref{eq:explicit_diffeo} is to be analyzed independently, beginning with the scalar $u$, i.e., by setting $s^{\mu \nu}=0$. The corresponding action is given by
\begin{subequations}
\begin{equation}
S_g|_{s^{\mu\nu}=0}=\int_{\mathcal{M}} \mathrm{d}^4x\,\frac{\sqrt{-g}}{2\kappa}(R+ \mathcal L'_{u})  \,,
\end{equation}
with 
 \begin{equation}
 \mathcal L'_{u}= -u   R \,.
 \end{equation}
\end{subequations}
The modified Einstein equations \cite{Bailey:2006fd} then read
\begin{subequations}
\begin{equation}
   G_{\mu\nu}-(T^{Ru})_{\mu\nu}=0\,,
\end{equation}
where
\begin{align}
\label{EMTu}
    (T^{Ru})_{\mu\nu}&=uG_{\mu\nu}-\frac 1 2 (\nabla_\mu \nabla_\nu u+ \nabla_\nu \nabla_\mu u) \notag \\
    &\phantom{{}={}}+g_{\mu\nu}\Box u\,,
\end{align}
\end{subequations}
with the d'Alembertian $\Box=\nabla_{\mu} \nabla^{\mu}$.
Our first step is to establish the identity~\eqref{eq:principal-equation-differential-form}. We have 
\begin{equation}
J:=\frac{\delta \mathcal L'_{u}}{\delta u}=-R \,.
\end{equation}
Since the covariant derivative of $u$ is the only quantity that can carry a free index, 
the right-hand side of Eq.~\eqref{eq:principal-equation-differential-form} divided by a factor of 2, reads
\begin{equation}
\label{RHS_identity}
\frac{J}{2} \nabla_\nu u=-\frac{R}{2}\nabla_\nu u\,.
\end{equation}
A direct evaluation of the left-hand side of Eq.~\eqref{eq:principal-equation-differential-form}, divided by 2, shows that
\begin{align}
    \nabla_\mu(T^{Ru})^{\mu}_{\phantom{\mu}\nu} &= G_{\mu\nu}\nabla^\mu u+ \big[\nabla_\nu,\nabla_\mu\big]
    \nabla^\mu u  \,.
\end{align}
By using 
\begin{equation}
\big[\nabla_\nu,\nabla_\mu\big]V^\mu=-R_{\nu\mu}V^\mu\,,
\end{equation}
for an arbitrary vector field $V^{\mu}$, we arrive at
\begin{equation}
\label{eq:consistency-requirement-u}
    \nabla_\mu(T^{Ru})^{\mu}_{\phantom{\mu}\nu}= G_{\mu\nu}\nabla^\mu u- 
    R_{\mu\nu}\nabla^\mu u=-\frac{R}{2}\nabla_\nu u\,,
\end{equation}
which, comparing with Eq.~\eqref{RHS_identity}, proves the identity. Now, consistency 
with Riemannian geometry dictates the validity of Eq.~\eqref{eq:consistency-requirement}. 
So we can either choose $\nabla_{\nu}u=0$, which amounts to a trivial constant background, or 
restrict spacetime geometry to pseudo-Riemannian 
manifolds with $R=0$; cf. Eq.~\eqref{eq:conditions-bailey} for nonzero $u$.

By taking the trace of the modified Einstein equations, we find an intriguing 
relationship between spacetime geometry and the scalar background:
\begin{equation}
-R=\frac{3}{1-u}\square u\,.
\end{equation}
The latter allows us to express $R$ in Eq.~\eqref{eq:consistency-requirement-u} in terms of $u$, which implies
\begin{equation}
\nabla_{\mu}(T^{Ru})^{\mu}_{\phantom{\mu}\nu}=\frac{3}{2(1-u)}\square u\nabla_{\nu}u=0\,.
\end{equation}
Demanding that $R=0$ is equivalent to considering backgrounds that obey
\begin{equation}
\label{eq:wave-equation-u}
\square u=0\,.
\end{equation}
Interestingly, this is a homogeneous wave equation for $u$, which allows for nontrivial solutions.

Now, we impose that the cosmological model based on Eq.~\eqref{eq:explicit_diffeo}, 
which implies explicit spacetime symmetry breaking, respects the symmetries dictated by the FLRW metric of Eq.~\eqref{metricFLRW}. For this to be the case, we demand that the object $(T^{Ru})^{\mu \nu}$ of Eq.~\eqref{EMTu} be form-invariant in the directions of the Killing vector fields.

A proper treatment of this problem requires the $(3+1)$ decomposition \cite{Arnowitt:1962hi} 
of the spacetime manifold $\mathcal{M}$. Let $e^{\mu} _a=e^{\mu} _a(x)$ be the tangent vector of a purely 
spacelike hypersurface of the spacetime foliation and $n^{\mu}=n^{\mu}(x)$ be the normal to this hypersurface. 
Using these variables allows us to decompose the tensor~\eqref{EMTu} into orthogonal components with respect 
to the foliation; see Refs.~\cite{Reyes:2024hqi,Reyes:2021cpx,Reyes:2022mvm}. Based on 
the FLRW metric of Eq.~\eqref{metricFLRW}, we arrive at%%
\begin{subequations}
\label{eq:decomposition-TRu}
\begin{align}
(T^{Ru})^{\mu\nu}&=e^\mu_a e^\nu_b T_{u}^{ab}+T_u^{a\mathbf{n}}(e^\mu_a n^\nu+n^\mu e^\nu_a) \notag \\[1ex]
    &\phantom{{}={}}+n^\mu n^\nu T_{u}^{\mathbf{n}\mathbf{n}}\,,
\end{align}
with the component
\begin{align}
   T_u^{ab}&=-\frac{1}{2}\bigg(D^aD^b u+D^bD^a u\bigg)+q^{ab}\bigg( D^2 u-2H\mathcal{L}_t u  \notag \\  
    &\phantom{{}={}}- \mathcal{L}_t^2 u  -\frac{k u}{a(t)^2}  -(2\dot{H}+3H^2)u\bigg)  \,,
\end{align}
being tangent to the hypersurfaces, the mixed contribution
\begin{equation}
T_u^{a\mathbf{n}}=D^a\big(\mathcal{L}_t u-H u\big)\,,
\end{equation}
and a part
\begin{equation}
    T_u^{\mathbf{n}\mathbf{n}}=-D^2 u+3H\mathcal{L}_t u+3\bigg(\frac{k}{a(t)^2}+H^2\bigg) u\,,
\end{equation}
defined in the one-dimensional space orthogonal to the hypersurfaces.
\end{subequations}
Here, $\mathcal L_t$ denotes the Lie derivative along the time flux four-vector and $D_a$ is the covariant derivative compatible with the induced metric $q_{ab}$, which is the spatial part of the FLRW metric of Eq.~\eqref{metricFLRW}. Furthermore, $q^{ab}$ denotes the corresponding inverse metric and $H=\dot a/ a$ is the Hubble parameter.

The tensor $(T^{Ru})^{\mu\nu}$ is form-invariant under the diffeomorphisms along each of the six Killing vector fields $\chi$ that describe isotropy and homogeneity of the FLRW metric. Then, the tangent-normal components $T_u^{a\mathbf{n}}$ must vanish and the spatial components have to be proportional to the induced metric of the $(3+1)$ decomposition~\cite{Reyes:2024hqi}.
Those conditions imply that the background satisfies
\begin{equation}
\mathcal L_{\chi}u(x)=0 \,,
\end{equation}
for each Killing vector field $\chi$. The only possibility of doing so is that the scalar background field depends on time only:
\begin{equation}
    u(x)\equiv u(t)\,.
\end{equation}
Then, according to Eq.~\eqref{eq:principal-equation} we have $\delta S_{\mathrm{obs}}=\delta S_{\mathrm{part}}$ in the Killing directions. For these generating vector fields, there is no clash between pseudo-Riemannian geometry and dynamics.

Finally, resorting to the wave equation for $u$, Eq.~\eqref{eq:wave-equation-u}, the latter can be recast for a purely time-dependent $u$ as follows:
\begin{equation}
\ddot{u}(t)+3H\dot{u}(t)=0\,,
\end{equation}
which provides a background field of the general form
\begin{equation}
\label{eq:background-u-solution}
u(t)=\alpha \int_0^t \mathrm{d}t'\frac{1}{a(t ')^3} +\beta\,,
\end{equation}
with the constants $\alpha=\dot{u}(0) a(0)^3$ and $\beta=u(0)$ depending on the initial conditions of the scale factor. Once a scale factor is obtained from the dynamics, the latter choice for $u$ is nontrivial and satisfies all requirements, in particular, Eq.~\eqref{eq:consistency-requirement}.

%...................................
\subsection{$s^{\mu \nu}$ sector}
%...................................

Next, we consider the $s^{\mu \nu}$ sector whose action is
\begin{subequations}
\begin{equation}
S_{g}|_{u=0}=\int_{\mathcal{M}} \mathrm{d}^4x\,\frac{\sqrt{-g}}{2\kappa}(R+\mathcal{L}'_{s})\,,
\end{equation}
with 
\begin{equation}
\mathcal{L}'_{s}=s^{\mu \nu}R_{\mu\nu}\,.
\end{equation}
\end{subequations}
The modified Einstein equations \cite{Bailey:2006fd} are given by
\begin{subequations}
\begin{equation}
G_{\mu\nu}-(T^{Rs})_{\mu\nu}=0\,,
\end{equation}
where
\begin{align}
\label{E-Ms}
    (T^{Rs})_{\mu\nu}&=\frac{1}{2}\Big(s^{\rho\sigma}R_{\rho\sigma}g_{\mu\nu}+g_{\nu\kappa}\nabla_\lambda \nabla_\mu s^{\lambda\kappa}\nonumber \\
    &\phantom{{}={}}\quad+g_{\mu\kappa}\nabla_\lambda\nabla_\nu s^{\lambda\kappa}-g_{\mu\lambda}g_{\nu\kappa}\Box s^{\lambda\kappa}\nonumber \\
    &\phantom{{}={}}\quad-g_{\mu\nu}\nabla_\rho\nabla_\sigma s^{\rho\sigma}\Big)\,.
\end{align}
\end{subequations}
Again, let us prove the identity of Eq.~\eqref{eq:principal-equation-differential-form} and consider 
\begin{equation}
J_{\alpha \beta}:= \frac{\delta\mathcal{L}'_{s}}{\delta s^{\alpha\beta}}= R_{\alpha \beta} \,.
\end{equation}
Then, the right-hand side of Eq.~\eqref{eq:principal-equation-differential-form} divided by 2, can be expressed as follows:
\begin{align}
\label{eq:right-hand-side-identity-s}
\frac{1}{2}\Big(J_{\alpha\beta}\nabla_{\nu}s^{\alpha\beta}&+\nabla_{\lambda}(J_{\nu\beta}s^{\lambda\beta})+\nabla_{\lambda}(J_{\alpha\nu}s^{\alpha\lambda})\Big) \notag \\
&=\frac{1}{2}R_{\alpha \beta} \nabla_{\nu}  s^{\alpha \beta}  +  \nabla_{\alpha}  (s^{\alpha \beta}  R_{\nu\beta}) \,.
\end{align}%%
To evaluate its left-hand side, we benefit from the identities 
\begin{subequations}
\begin{align}
\label{id1}
[\nabla _{\alpha},\nabla _{\beta}  ] T^{\mu_1 \mu_2\dots \mu_n}&=R^{\mu_1}_{\phantom{\mu_1}\rho \alpha \beta} T^{\rho \mu_2\dots \mu_n}+ R^{\mu_2}_{\phantom{\mu_2} \rho \alpha \beta} T^{ \mu_1 \rho \dots \mu_n}\notag \\
&\phantom{{}={}}+\dots +R^{\mu_n}_{\phantom{\mu_n} \rho \alpha \beta} T^{ \mu_1 \dots  \rho} \,,
\end{align}
for a generic contravariant tensor $T^{\mu_1 \mu_2\dots \mu_n}$ of rank $n$ and
\begin{equation}
\label{id2}
R_{  \alpha \beta; \rho}  -R_{  \alpha  \rho;\beta}  +R^{\mu}_{\phantom{\mu}\alpha \beta \rho; \mu}  =0   \,,
\end{equation}
for covariant derivatives of the Riemann and Ricci tensor, respectively.
\end{subequations}
It is then possible to show that
\begin{align}
\nabla_{\mu}(T^{Rs})^{\mu}_{\phantom{\mu}\nu}&=\frac{1}{2}R_{\alpha\beta} \nabla_{\nu} s^{\alpha\beta}  + \nabla_{\alpha}  (s^{\alpha\beta}  R_{\nu\beta})  \,.
\end{align}%%
The latter agrees with Eq.~\eqref{eq:right-hand-side-identity-s}, whereupon Eq.~\eqref{eq:principal-equation-differential-form} has been demonstrated successfully.

By taking into account the isometries along the Killing vector fields $\chi$, the background field must obey
\begin{equation}
  \mathcal L_{\chi} s^{\mu \nu}=0 \,.
\end{equation}
For Killing directions associated with homogeneity and isotropy, this requirement is satisfied by the generic choice
\begin{equation}
\label{eq:s-sector}
    s^{\mu\nu}(x)=e^\mu_a e^\nu_b s_1(t)q^{ab}+n^\mu n^\nu s_2(t)\,,
\end{equation}
where $s_1(t)$ and $s_2(t)$ are arbitrary functions of time. As before in Eq.~\eqref{eq:decomposition-TRu}, we benefit from the variables $e^{\mu}_a$ and $n^{\mu}$ employed in the $(3+1)$ decomposition of spacetime~\cite{Arnowitt:1962hi}. By inserting Eq.~\eqref{eq:s-sector} into Eq.~\eqref{E-Ms}, we obtain the expression
\begin{subequations}
\begin{equation}
    (T^{Rs})^{\mu\nu}=e^\mu_{a}e^\nu_{b}T_s^{ab}+n^\mu n^\nu T_s^{\mathbf{n} \mathbf{n} }\,,
\end{equation}
with
\begin{align}
    T_s^{ab}&=\bigg[\bigg(\frac{3k}{a(t)^2}+\dot{H}+3H^2\bigg)s_1(t)-(2\dot{H}+3H^2)s_2(t) \notag \\
    &\phantom{{}={}}\quad-2H\dot{s}_2(t)+\frac{1}{2}\big(\ddot{s}_1(t)-\ddot{s}_2(t)\big)\bigg]q^{ab}\,,
\end{align}
and
\begin{align}
\label{eq:T_nn}
    T_s^{\mathbf{n} \mathbf{n} }&=-\frac{3k}{a(t)^2}s_1(t)+3(\dot{H}+2H^2)s_2(t)\notag \\
    &\phantom{{}={}}-\frac{3}{2}H\big(\dot{s}_1(t)-\dot{s}_2(t)\big)\,,
\end{align}
\end{subequations}
where only the time dependencies of $a$, $s_1$, and $s_2$ are kept, for brevity.
Finally, requiring that the object $(T^{Rs})^{\mu\nu}$ be divergence-free according to Eq.~\eqref{eq:consistency-requirement}
we arrive at a differential equation for the scale factor and the time-dependent functions in Eq.~\eqref{eq:s-sector}:
\begin{align}
     0&= \bigg(\frac{2k}{a(t)^2} + \dot{H}+3H^2\bigg)\big[  \dot{s}_1(t)-2Hs_1(t)\big]\notag \\ 
     &\phantom{{}={}}-2s_2(t)  (2H\dot{H}+\ddot{H})\notag \\
    &\phantom{{}={}}-3(H^2+\dot{H}) \big[    \dot{s}_2(t)+2Hs_2(t)\big]\,. 
    \label{sbianchi}
\end{align}
What remains is to find nontrivial solutions of the previous equation. In what follows, we will analyze two possible
background field configurations.

%...................................
\subsubsection{Traceless configuration}
%...................................

First, let the background field be traceless:
\begin{equation}
    g_{\mu\nu}s^{\mu\nu}=0 \,,
\end{equation}
which, from Eq.~\eqref{eq:s-sector}, implies
\begin{equation}
    3s_1(t)=s_2(t)\,.
\end{equation}
By taking this into account, one can show that one possible solution of Eq.~\eqref{sbianchi} is of the form
\begin{subequations}
\label{eq:background-s-solution-1}
\begin{equation}
    (s^{\mu\nu})=\left(\begin{matrix}
        3s_1(t) & 0 & 0 & 0 \\
        0 & q^{rr}s_1(t) & 0 & 0 \\
        0&0&q^{\theta\theta}s_1(t)&0 \\
        0&0&0&q^{\phi\phi}s_1(t)\end{matrix}\right)\,,
\end{equation}
where, explicitly,
\begin{align}
    s_1(t)&=s_1(0)\exp\Bigg[\int_0^t \mathrm{d}t'\,\Xi(t')\Bigg] \,, \\[1ex]
    \Xi(t')&=\frac{2\Big(\frac{k}{a(t')^2} +8\dot{H}+6H^2\Big)H +3\ddot{H}}{\frac{k}{a(t')^2} -4\dot{H} -3H^2}\,.
\end{align}
\end{subequations}
Here, the purely spacelike components of the metric inverse to Eq.~\eqref{metricFLRW} are needed, as well, which read
\begin{subequations}
\label{eq:inv_metric}
\begin{align}
q^{rr}&=\frac{1-kr^2}{a(t)^2} \,, \displaybreak[0]\\
q^{\theta\theta}&=\frac{1}{r^2 a(t)^2} \,, \displaybreak[0]\\
q^{\phi\phi}&=\frac{1}{r^2 \sin^2\theta a(t)^2}\,.
\end{align}
\end{subequations}
With dynamics giving rise to a scale factor, Eq.~\eqref{eq:background-s-solution-1} can be stated explicitly. It poses a nontrivial nondynamical background field that does not clash with Riemannian geometry.

%...................................
\subsubsection{Configuration with nonzero trace}
%...................................

Next, we study a configuration for the background field with different properties. An observation from Eq.~\eqref{sbianchi} is that each term containing the time derivatives of $s_1(t)$ and $s_2(t)$, respectively, appears with a global function scaled by $H$. This suggests choosing a background configuration satisfying
\begin{subequations}
\begin{align}
\label{eq:s1ansatz}
    \dot{s}_1(t)&=\alpha H s_1(t)\,, \\[1ex]
\label{eq:s2ansatz}
    \dot{s}_2(t)&=\beta H s_2(t)\,,
\end{align}
\end{subequations}
with constant parameters $\alpha$ and $\beta$. By inserting the latter into Eq.~\eqref{sbianchi}, we arrive at the differential equation
\begin{align}
0&=\bigg(\frac{2k}{a(t)^2} + \dot{H}+3H^2\bigg)  (\alpha-2)Hs_1(t)\notag \\
&\phantom{{}={}}-2\frac{s_2(t)}{a(t)}   \Bigg\{\partial_t\big[a(t)(\dot{H}+H^2)\big]\notag \\
&\phantom{{}={}}+\bigg(\frac{3}{2}\beta+2\bigg)(H^2+\dot{H})\dot{a}(t)\Bigg\}\,.
\end{align}
The intriguing choices $\alpha=2$ and $\beta=-4/3$ lead to a quite simple equation, which is independent of the functions composing the background field:
\begin{equation}
   \frac{\partial}{\partial t}\big[a(t)(\dot{H}+H^2)\big]=0 \,.
\end{equation}
The latter has the solution $\ddot{a}=\text{const.}$, which is to be considered. By solving Eqs.~\eqref{eq:s1ansatz} and \eqref{eq:s2ansatz}, we are able to state an alternative to Eq.~\eqref{eq:background-s-solution-1}:
\begin{widetext}
\begin{equation}
\label{eq:background-s-solution-2}
    (s^{\mu\nu})=\frac{a(t)^2}{a(0)^2}\left(\begin{matrix}
    s_2(0)\frac{a(0)^{10/3}}{a(t)^{10/3}} & 0 & 0 & 0 \\ 
    0 & q^{rr}s_1(0) & 0 & 0 \\
    0&0&q^{\theta\theta}s_1(0)&0 \\
    0&0&0&q^{\phi\phi}s_1(0)
    \end{matrix}\right)\,,
\end{equation}
\end{widetext}
with the components of the inverse FLRW metric given by Eq.~\eqref{eq:inv_metric}. Again, the latter is a nontrivial background field that is in accordance with the symmetries demanded and satisfies Eq.~\eqref{eq:consistency-requirement}. Consequently, it does not provide a discrepancy with the second Bianchi identities of Riemannian geometry.%

Note that the modified Friedmann equations alone, which govern the dynamics, cannot fix the background field. To do so, the second Bianchi identities must be imposed. By inserting Eqs.~\eqref{eq:background-u-solution}, \eqref{eq:background-s-solution-1} or \eqref{eq:background-s-solution-2} into the corresponding dynamical equations, they become integro-differential equations for the scale factor, which are, in principle, solvable numerically. Ultimately, the system is fully determined and the background field can be stated as a function of time; see Ref.~\cite{Reyes:2024hqi}.

%............................
\section{Conclusions}
\label{sec:conclusions}
%............................

In this paper, some of the essential aspects of explicit diffeomorphism violation in gravity were reconsidered. First, we explained in a transparent way the possible inconsistency between dynamics and spacetime geometry through Eqs.~\eqref{eq:principal-equation} and \eqref{eq:principal-equation-differential-form}. Second, the approach of restricting spacetime geometry in the cases of the gravitational CS term~\cite{Jackiw:2003pm} and the minimal gravitational SME~\cite{Bailey:2024zgr} was revisited. For consistency, one must impose spacetime geometries with ${}^{*}RR=0$ for the CS term and $R=0$ for a subset of the $u$, $s$, and $t$ coefficients.
Third, we introduced an alternative approach, which substantially relies on isometries of a gravitational system. As shown in Ref.~\cite{Reyes:2024hqi} in the specific setting of the $t$ coefficients in a FLRW spacetime, it allows for a treatment of explicit diffeomorphism violation in gravity.

We have identified further examples of consistent models with explicit diffeomorphism breaking governed by the SME background fields $u$ and $s^{\mu\nu}$ in the context of cosmology. Both models have been shown to be consistent with the Bianchi identities of pseudo-Riemannian geometry. They also respect the principles of isotropy and homogeneity. Therefore, this paper complements the examples encountered in the contemporary literature. 
This technique is expected to be applicable to other gravitational systems such as black holes, which are also accompanied by isometries.

Bailey \textit{et al.}~\cite{Bailey:2024zgr} showed recently how linearized-gravity theories with explicit diffeomorphism violation can imply more than 2 propagating physical degrees of freedom, where the additional ones are not suppressed by SME coefficients. This poses an excellent opportunity to look for such modes in gravitational waves and to rule out these extensions if additional modes are not detected. 

Our approach relies on the full nonlinear modified-gravity theory defined by the action of the gravitational SME. It demonstrates how isometries can be valuable for constructing nontrivial nondynamical background fields that do not clash with Riemannian geometry, at least for a subset of metrics compatible with the symmetries of the system. These results form a robust theoretical base for studying the phenomenology of such settings further in the future.

\section*{Acknowledgments}

We thank J. Belinchon for bringing to our attention the different kinds of continuous symmetries that were recapitulated in Sec.~\ref{sec:introduction}. We appreciate that the referee encouraged us to support our results by additional explicit examples. CMR acknowledges partial support to the Fondecyt Regular 1241369 project and is grateful to M.~Schreck and M.M.~Ferreira, Jr. for their kind hospitality as well as the postgraduate program of the Physics Department at UFMA where this work was conceived. CR acknowledges support from the ANID fellowship No. 21211384 and Universidad de Concepci\'on. MS is indebted to CNPq Produtividade 310076/2021-8 and CAPES/Finance Code 001.

%..........................................................................

%\bibitem{a}
%Author, \emph{Title}, \emph{J. Abbrev.} {\bf vol} (year) pg.
%
%\bibitem{b}
%Author, \emph{Title},
%arxiv:1234.5678.
%
%\bibitem{c}
%Author, \emph{Title},
%Publisher (year).
%
%
%% Please avoid comments such as "For a review'', "For some examples",
%% "and references therein" or move them in the text. In general,
%% please leave only references in the bibliography and move all
%% accessory text in footnotes.
%
%% Also, please have only one work for each \bibitem.
%
%


\begin{thebibliography}{99}
%..........................................................................

%\cite{Blau:2020}
\bibitem{Blau:2020}
M.~Blau,
{\em Lecture Notes on General Relativity},
lecture notes (Bern University, 2002), \url{http://www.blau.itp.unibe.ch/GRLecturenotes.html}.

%\cite{Bluhm:2014oua}
\bibitem{Bluhm:2014oua}
R.~Bluhm,
``Explicit versus spontaneous diffeomorphism breaking in gravity,''
Phys. Rev. D \textbf{91}, 065034 (2015),
%doi:10.1103/PhysRevD.91.065034
arXiv:1401.4515 [gr-qc].

%\cite{Kostelecky:1988zi}
\bibitem{Kostelecky:1988zi}
V.A.~Kosteleck\'{y} and S.~Samuel,
``Spontaneous breaking of Lorentz symmetry in string theory,''
Phys. Rev. D \textbf{39}, 683 (1989).

%\cite{Kostelecky:1989jp}
\bibitem{Kostelecky:1989jp}
V.A.~Kosteleck\'{y} and S.~Samuel,
``Phenomenological gravitational constraints on strings and higher-dimensional theories,''
Phys. Rev. Lett. \textbf{63}, 224 (1989).

%\cite{Kostelecky:1989jw}
\bibitem{Kostelecky:1989jw}
V.A.~Kosteleck\'{y} and S.~Samuel,
``Gravitational phenomenology in higher-dimensional theories and strings,''
Phys. Rev. D \textbf{40}, 1886 (1989).

%\cite{Kostelecky:1991ak}
\bibitem{Kostelecky:1991ak}
V.A.~Kosteleck\'{y} and R.~Potting,
``{\em CPT} and strings,''
Nucl. Phys. B \textbf{359}, 545 (1991).

%\cite{Kostelecky:1994rn}
\bibitem{Kostelecky:1994rn}
V.A.~Kosteleck\'{y} and R.~Potting,
``{\em CPT}, strings, and meson factories,''
Phys. Rev. D \textbf{51}, 3923 (1995),
arXiv:hep-ph/9501341.

%\cite{Kostelecky:2003fs}
\bibitem{Kostelecky:2003fs}
V.A.~Kosteleck\'{y},
``Gravity, Lorentz violation, and the standard model,''
Phys. Rev. D \textbf{69}, 105009 (2004),
arXiv:hep-th/0312310.
%1016 citations counted in INSPIRE as of 21 Apr 2020

%\cite{Kostelecky:2020hbb}
\bibitem{Kostelecky:2020hbb}
V.A.~Kosteleck\'{y} and Z.~Li,
``Backgrounds in gravitational effective field theory,''
Phys. Rev. D \textbf{103}, 024059 (2021),
arXiv:2008.12206 [gr-qc].

%\cite{Bluhm:2016dzm}
\bibitem{Bluhm:2016dzm}
R.~Bluhm and A.~Sehic,
``Noether identities in gravity theories with nondynamical backgrounds and explicit spacetime symmetry breaking,''
Phys. Rev. D \textbf{94}, 104034 (2016),
%doi:10.1103/PhysRevD.94.104034
arXiv:1610.02892 [hep-th].

%\cite{Bluhm:2004ep}
\bibitem{Bluhm:2004ep}
R.~Bluhm and V.A.~Kosteleck\'{y},
``Spontaneous Lorentz violation, Nambu-Goldstone modes, and gravity,''
Phys. Rev. D \textbf{71}, 065008 (2005),
%doi:10.1103/PhysRevD.71.065008
arXiv:hep-th/0412320.

%\cite{Bluhm:2007bd}
\bibitem{Bluhm:2007bd}
R.~Bluhm, S.H.~Fung, and V.A.~Kosteleck\'{y},
``Spontaneous Lorentz and diffeomorphism violation, massive modes, and gravity,''
Phys. Rev. D \textbf{77}, 065020 (2008),
%doi:10.1103/PhysRevD.77.065020
arXiv:0712.4119 [hep-th].

%\cite{Kostelecky:2010hs}
\bibitem{Kostelecky:2010hs}
V.A.~Kosteleck\'{y} and N.~Russell,
``Classical kinematics for Lorentz violation,''
Phys. Lett. B \textbf{693}, 443 (2010),
%doi:10.1016/j.physletb.2010.08.069
arXiv:1008.5062 [hep-ph].

%\cite{Kostelecky:2011qz}
\bibitem{Kostelecky:2011qz}
V.A.~Kosteleck\'{y},
``Riemann-Finsler geometry and Lorentz-violating kinematics,''
Phys. Lett. B \textbf{701}, 137 (2011),
%doi:10.1016/j.physletb.2011.05.041
arXiv:1104.5488 [hep-th].

%\cite{AlanKostelecky:2012yjr}
\bibitem{AlanKostelecky:2012yjr}
V.A. Kosteleck\'y, N.~Russell, and R.~Tso,
``Bipartite Riemann-Finsler geometry and Lorentz violation,''
Phys. Lett. B \textbf{716}, 470 (2012),
%doi:10.1016/j.physletb.2012.09.002
arXiv:1209.0750 [hep-th].

%\cite{Schreck:2015seb}
\bibitem{Schreck:2015seb}
M.~Schreck,
``Classical Lagrangians and Finsler structures for the nonminimal fermion sector of the Standard-Model Extension,''
Phys. Rev. D \textbf{93}, 105017 (2016),
%doi:10.1103/PhysRevD.93.105017
arXiv:1512.04299 [hep-th].

%\cite{Stueckelberg:1938}
\bibitem{Stueckelberg:1938}
E.C.G.~Stueckelberg,
``Die Wechselwirkungskr\"{a}fte in der Elektrodynamik und in der Feldtheorie der Kernkr\"{a}fte. Teil II und III.,
Helv.\ Phys.\ Acta \textbf{11}, 299 (1938).

%\cite{Ruegg:2003ps}
\bibitem{Ruegg:2003ps}
H.~Ruegg and M.~Ruiz-Altaba,
``The Stueckelberg field,''
Int. J. Mod. Phys. A \textbf{19}, 3265 (2004),
%doi:10.1142/S0217751X04019755
arXiv:hep-th/0304245.

%\cite{Arkani-Hamed:2002bjr}
\bibitem{Arkani-Hamed:2002bjr}
N.~Arkani-Hamed, H.~Georgi, and M.D.~Schwartz,
``Effective field theory for massive gravitons and gravity in theory space,''
Annals Phys. \textbf{305}, 96 (2003),
%doi:10.1016/S0003-4916(03)00068-X
arXiv:hep-th/0210184.

%\cite{Hinterbichler:2011tt}
\bibitem{Hinterbichler:2011tt}
K.~Hinterbichler,
``Theoretical aspects of massive gravity,''
Rev. Mod. Phys. \textbf{84}, 671 (2012),
%doi:10.1103/RevModPhys.84.671
arXiv:1105.3735 [hep-th].

%\cite{Bluhm:2019ato}
\bibitem{Bluhm:2019ato}
R.~Bluhm, H.~Bossi, and Y.~Wen,
``Gravity with explicit spacetime symmetry breaking and the Standard-Model Extension,''
Phys. Rev. D \textbf{100}, 084022 (2019),
%doi:10.1103/PhysRevD.100.084022
arXiv:1907.13209 [gr-qc].

%\cite{Jackiw:2003pm}
\bibitem{Jackiw:2003pm}
R.~Jackiw and S.-Y.~Pi,
``Chern-Simons modification of general relativity,''
Phys. Rev. D \textbf{68}, 104012 (2003),
%doi:10.1103/PhysRevD.68.104012
arXiv:gr-qc/0308071.

%\cite{Obukhov:1995eq}
\bibitem{Obukhov:1995eq}
Y.N.~Obukhov and F.W.~Hehl,
``On the relation between quadratic and linear curvature Lagrangians in Poincar\'{e} gauge gravity,''
Acta Phys. Polon. B \textbf{27}, 2685 (1996),
arXiv:gr-qc/9602014.

%\cite{Bailey:2024zgr}
\bibitem{Bailey:2024zgr}
Q.G.~Bailey, K.~O'Neal-Ault, and N.A.~Nilsson,
``Explicit diffeomorphism violation no-go constraints and discontinuities,''
Phys. Rev. D \textbf{110}, 084066 (2024),
%doi:10.1103/PhysRevD.110.084066
arXiv:2407.04918 [gr-qc].

%\cite{Reyes:2024hqi}
\bibitem{Reyes:2024hqi}
C.M.~Reyes, C.~Riquelme, and A.~Soto,
``Cosmology with explicit and spontaneous background fields,''
JCAP \textbf{05}, 014 (2025),
%doi:10.1088/1475-7516/2025/05/014
arXiv:2407.13041 [gr-qc].

%\cite{York:1972sj}
\bibitem{York:1972sj}
J.W.~York, Jr.,
``Role of Conformal Three-Geometry in the Dynamics of Gravitation,''
Phys. Rev. Lett. \textbf{28}, 1082 (1972).

%\cite{Gibbons:1976ue}
\bibitem{Gibbons:1976ue}
G.W.~Gibbons and S.W.~Hawking,
``Action integrals and partition functions in quantum gravity,''
Phys. Rev. D \textbf{15}, 2752 (1977).

\bibitem{Reyes:2021cpx}
C.M.~Reyes and M.~Schreck,
``Hamiltonian formulation of an effective modified gravity with nondynamical background fields,''
Phys. Rev. D \textbf{104}, 124042 (2021).

\bibitem{Reyes:2022mvm}
C.M.~Reyes and M.~Schreck,
``Modified-gravity theories with nondynamical background fields,''
Phys. Rev. D \textbf{106}, 044050 (2022).

%\cite{Reyes:2023sgk}
\bibitem{Reyes:2023sgk}
C.M.~Reyes and M.~Schreck,
``The boundary of the gravitational standard-model extension,''
Phys. Rev. D \textbf{108}, 104013 (2023),
%doi:10.1103/PhysRevD.108.104013
arXiv:2305.07060 [gr-qc].

%\cite{Yano:1957}
\bibitem{Yano:1957}
K.~Yano,
\textit{Lie Derivatives and its Applications}
(North-Holland Publishing Co., Amsterdam; P. Noordhoff Ltd., Groningen, The Netherlands, 1957 --- Reprinted by Dover Publications, Inc., Mineola, NY (USA), 2020).

%\cite{Carroll:1997ar}
\bibitem{Carroll:1997ar}
S.M.~Carroll,
\textit{Lecture notes on general relativity},
arXiv:gr-qc/9712019.

%\cite{xTensor:2024}
\bibitem{xTensor:2024}
J.M.~Mart\'{i}n-Garc\'{i}a,
\textit{xAct:  Efficient tensor computer algebra for the Wolfram Language},
\url{http://www.xact.es}.

%\cite{Bailey:2006fd}
\bibitem{Bailey:2006fd}
Q.G.~Bailey and V.A.~Kosteleck\'{y},
``Signals for Lorentz violation in post-Newtonian gravity,''
Phys. Rev. D \textbf{74}, 045001 (2006),
%doi:10.1103/PhysRevD.74.045001
arXiv:gr-qc/0603030.

%\cite{Arnowitt:1962hi}
\bibitem{Arnowitt:1962hi}
R.L.~Arnowitt, S.~Deser, and C.W.~Misner,
\textit{The dynamics of general relativity},
in {\em Gravitation: An Introduction to Current Research}, L. Witten (ed.)
(Wiley, New York, 1962).

\end{thebibliography}
\end{document}